\documentclass{article}
\pdfoutput=1
\usepackage{arxiv}

\usepackage{url}
\usepackage{graphicx}
\usepackage{wrapfig}
\usepackage{xcolor}

\PassOptionsToPackage{hyphens}{url} 

\begin{document}
\title{MEDFORD in a Box: Improvements and Future Directions for a Metadata Description Language}

\author{
Polina Shpilker \\
Department of Computer Science \\
Tufts University \\
Medford, MA \\
\And
Benjamin Stubbs \\
Department of Computer Science \\
Tufts University \\
Medford, MA \\
\And
Michael Sayers \\
Department of Computer Science and Statistics\\
University of Rhode Island\\
Kingston, RI\\
\And
Yumin Lee \\
Department of Computer Science \\
Tufts University \\
Medford, MA \\
\And
Lenore Cowen \\
Department of Computer Science \\
Tufts University \\
Medford, MA \\
\And
Donna Slonim \\
Department of Computer Science \\
Tufts University \\
Medford, MA \\
\And
Shaun Wallace \\
Department of Computer Science and Statistics\\
University of Rhode Island\\
Kingston, RI\\
\texttt{shaun.wallace@uri.edu} \\
\And
Alva Couch \\
Department of Computer Science\\
Tufts University \\
Medford, MA \\
\texttt{alva.couch@tufts.edu} \\
\And
Noah M. Daniels \\
Department of Computer Science and Statistics\\
University of Rhode Island\\
Kingston, RI\\
\texttt{noah\_daniels@uri.edu} \\
}



\maketitle

\begin{abstract}
Scientific research metadata is vital to ensure the validity, reusability,
and cost-effectiveness of research efforts. The MEDFORD metadata
language was previously introduced to simplify the process of writing
and maintaining metadata for non-programmers. However, barriers to
entry and usability remain, including
limited automatic validation, difficulty of  data transport,
and user unfamiliarity with text file editing. To address these
issues, we introduce MEDFORD-in-a-Box (MIAB), a documentation ecosystem
to facilitate researcher adoption and earlier metadata capture. MIAB contains
many improvements, including an updated MEDFORD parser with
expanded validation routines and BagIt export capability.
MIAB also includes an improved VS Code extension that supports these changes through a visual IDE. By simplifying metadata generation, this new tool supports the creation of correct, consistent, and reusable metadata, ultimately improving research reproducibility.

\end{abstract}
\section{Introduction}

Algorithms in the AI age are hungry not only for data, but also for data about data to provide contextual information. While many reasonable and successful FAIR (Findable, Accessible, Interoperable, and Reusable) metadata standards have been proposed~\cite{ball2016rda,qin2012functional}, using them in production is often a second thought, put off until research is complete and funding or publication rules require documentation. We believe that if it were easier to maintain metadata, especially in a findable way, more and better documentation would be generated earlier.

One possible existing solution for metadata maintenance is the Resource Description Framework (RDF), originally developed by the World Wide Web Consortium for this purpose~\cite{candan2001resource,manola2004rdf,pan2009resource}. RDF posed a more human-readable alternative to more machine-readable XML metadata formats~\cite{kaykova2005rscdf}. It is worth noting that RDF itself is not a language, but a framework. Multiple languages enable the production of RDF-compliant metadata, typically comprising a graph structure that relies on RDF's triples (entity, relationship, entity) to define edges within that graph. Such languages include Turtle~\cite{beckett2014rdf}, N3~\cite{berners2008n3logic}, RPV~\cite{brayrpv}, and Triple~\cite{sintek2002triple}.
However, while these higher-level languages present as human-readable, they are not designed to be used by scholars with no training in programming or RDF structure.

The MEtaData Format for Open Reef Data (MEDFORD)\cite{shpilker2022medford,shpilker2022metadata} language is a 
previously-proposed metadata description language specifically designed to help non-computational scientists describe their research processes and data in a FAIR manner. MEDFORD provides an easy-to-understand vocabulary that is extensible to meet the needs of users and metadata schemas that improve reusability and data quality. MEDFORD produces plain text-formatted files that are easily shareable and indexable, increasing findability, accessibility, and interoperability with other systems.

Our contributions here are twofold. First, we present a partial implementation of the proposed extension of the MEDFORD language presented in ~\cite{shpilker2024cross,sfakakis2025metadata}, with associated updates to the proposed syntax for macros and external references. Second, we present all the tools needed to easily write and check MEDFORD metadata files centered in Visual Studio Code, in a set of open sources packages we call MEDFORD-in-a-Box (MIAB). MIAB consists of an updated MEDFORD parser, a Visual Studio Code extension (first described in ~\cite{Strand2024-np}), the MEDFORD
language server protocol module, an extensive testing suite, data validation resources, external references, and also discusses an upcoming update to the proposed syntax for macros. We believe that MIAB will allow even non-programmers to confidently write MEDFORD metadata files.

\section{Extensions to the MEDFORD parser}

The MEDFORD parser is the workhorse of the MEDFORD system. It checks MEDFORD syntax, validates MEDFORD files, and creates archives for sharing. Beyond reporting language errors, the MEDFORD parser can learn the required attributes for various export file types and help ensure that the researcher provides these attributes. For example, NSF-funded coral research must be submitted to BCO-DMO\cite{bco-dmo}, which requires metadata such as @Contributor or @Cruise information.

The MEDFORD parser can help ensure that the provided metadata is in the correct format. In the newest version, MEDFORD provides validation routines for a number of basic major token data types, such as email, number, and text. These data types are combined with basic operations and a comparison value to allow for common data validation tasks to be defined easily, such as Version being a number $>$0 or that email addresses appear reasonable.

These rules are designed to be formulaic, allowing us to provide a method in the future for users to specify their own custom major tokens and validation routines without needing to write Python methods. We envision users defining tokens and validation rules in a simplified text format, such as YAML.

For example, consider the YAML describing a new entity ``Institution" where we define a few minor tokens such as address, city, and URI along with validation rules for them:

\begin{verbatim}
  Institution:
- Type: string
- Required # The tag is required for this output mode.
   It must be present or produce an error
- Multiple # there can be multiple instances of this tag
- Contents:
 - address: # minor token
  - Type: string # here type represents a validation subroutine
   called by the validator
  - Desirable # This means we want the minor tag to be present,
   but warn do not error
 - city:
  - Type: string
  - Desirable
 - province:
  - Desirable
 - country:
  - Type: string
 - URI:
  - Type: URI
  - Desirable
 - phone:
  - Type: phone
  - Desirable

\end{verbatim}

These YAML files could then be imported to provide additional data models. These new data models can also be paired with custom validation subroutines by advanced users if they follow the IO templates and link them in the MEDFORD validation mapping file MEDFORD.mvd.

The MEDFORD language was also built with transport in mind; using a technology called BagIt~\cite{kunze2018bagit}, the MEDFORD parser can completely envelop the MEDFORD metadata file alongside all referenced research data files into a single zip file as described by the BagIt format. This zip file contains hashes to ensure files are transferred without error and can denote related references, which may be remote. Specifically, \texttt{\_Primary} and \texttt{\_Copy} tags are stored within the zip file and their hashes are in the manifest file, but references mentioned with the \texttt{\_Ref} tag are neither stored in the zip file nor have their hashes in the manifest file. This allows a MEDFORD file to be transferred alongside the research data it describes, enabling a researcher to provide a complete description of of the data to a collaborator or submission database with a single file transfer. It also preserves the flexibility to refer to external data sources that may be too large to realistically transfer in the BagIt zip file, such as genomic datasets.

The newest version of the MEDFORD parser contains updated BagIT utilities focused on data quality and completeness. BagIT archives now have their own validation routines where every included file is checked to make sure it has an associated tag and all expected files are verified to be present based on the MEDFORD file before archiving.

\section{Visual Studio Code Extension Updates}

Visual Studio Code~\cite{vscode} (VS Code) is an industry-standard integrated development environment (IDE) with a thriving online community aimed at extending the capabilities of the editor through external software packages called extensions. While MEDFORD has had an extension available in the VS Code marketplace since 2022 \cite{Strand2024-np}, MIAB represents a major update and shift in how we are moving forward.

Microsoft pioneered the Language Server Protocol (LSP) as a way to make supporting language features, such as linting, syntax highlighting, and completion suggestions, portable across many different editors (technically, given $M$ editors and $N$ languages, LSP turns an $M\times N$ problem into an $M + N$ problem). This is because instead of each editor needing $N$ language plugins ($M \times N$), there need only be $M$ LSP plugins and $N$ language servers. Note that for some editors, an additional lightweight plugin may be desired for a language in order to support additional features, such as unusual build commands or formatting commands.

The VS Code extension is designed to be an easy entry point into using MEDFORD. Having installed VS Code, a user can install the MEDFORD parser, the MEDFORD language server, and the extension itself in only a few commands, and immediately start working with MEDFORD. The extension associates MEDFORD files with the filetype \texttt{.mfd} for easy opening. The extension provides syntax highlighting of the MEDFORD language to improve readability and decrease errors. While the extension is active, the language server runs in the background, providing helpful interactive feedback such as tab completions for minor tags.

This new release has been updated to use the most recent versions of pydantic and pyglas, refactored to streamline updates and reusability, and integrated with the new version of the MEDFORD parser to support new data models, validations, and export modes as they are released.

We have decoupled the MEDFORD parser, bagIT utilities, and language server components of the VS Code extension to enable independent development and reduce code duplication. For example, the language server is not intrinsically tied to the VS Code application, and by separating it, we allow for it to be integrated into other technologies, such as web-based IDEs like Monaco, or other editors that support the Language Server Protocol, such as neovim or Zed. By removing data model information from the extension, we have one point of truth for data models: the parser. This enables us to develop and release new models easily, promoting the future feature of importable user data models and validation tasks described above.

\section{Extensions to the MEDFORD Language Specification}

We have also upgraded the formal specifications for the MEDFORD
language to allow for the desired usability. Several of the key issues and extensions are detailed in this section.


\subsection{Metadata Inter-connectivity}
%

We initially designed MEDFORD to have a self-contained block structure. A MEDFORD block represented an entity, and all metadata about that entity was contained within it. Originally, there was no concept of context or relationships between entities. However, this quickly proved insufficient to describe organic research data.


Our initial user testing involved a group of coral researchers who wanted to use MEDFORD to document their research samples. The basic information needed to keep track of a coral species includes
its scientific name, the current (at the time of research) genome construction,
and the reef in which the species was studied. Details about these coral samples can be described in MEDFORD as follows:


\begin{verbatim}
  @Species P.Dam
  @Species_Reef New Caledonia Barrier reef
  @Species_Reef-Coordinates (coordinates)
\end{verbatim}

These researchers also needed to document the reefs in which the samples were collected. Reefs typically required information such as geo-coordinates, the date of study, and the coral species present in the reef, leading to a MEDFORD representation such as:

\begin{verbatim}
  @reef New Caledonia Barrier reef
  @reef-Coordinates (coordinates)
  @reef-coral_species Staghorn coral
  @reef-coral_species Elkhorn coral
\end{verbatim}

But, there is a strong dependency between the description of a coral species and a
reef: to completely describe one; there must be some description of the other.
This issue could have been resolved by simply making one of the major tokens
a minor token of the other token.


\begin{verbatim}
  @Species P.Dam
  @Species_Reef New Caledonia Barrier Reef
  @Species_Reef-Coordinates (coordinates)

  @Species P.Acuta
  @Species_Reef New Caledonia Barrier Reef
  @Species_Reef-Coordinates (coordinates, with a typo)
\end{verbatim}

But this leads to further complications. Here, every instance of a species from the same reef will have duplicated reef metadata. This duplication adds work, increases transcription errors, and imposes a hierarchy on the data that is collected. We are considering species 
to be the primary entity, which is a valid but arbitrary choice. If the researchers are focusing on the differences between {\em reefs}, however, they might prefer if @Species were a child token of \texttt{@Reef}.
This example demonstrates that we need a concept of relationships between metadata objects that is neither repetitive nor hierarchical.

\subsection{Naming Metadata Objects}


Originally, MEDFORD was not designed with a concept of unique identifiers, and without them, connecting entities is not practical. However, the first line is colloquially called the ``description'' or \texttt{desc} line of the metadata object in MEDFORD documentation.

\begin{verbatim}
  @Species P. Dam <-- this is the description
  @Species-Construction Pocillopora damicornis genome v1.0
\end{verbatim}

 This line was not utilized by the MEDFORD parser and typically served the user as a one-line description of the metadata object described by the following minor token lines, similar to the functionality of the \texttt{note} minor token, which allows users to create custom non-processed annotations for a given metadata object. Every block must have a \texttt{desc} line for the file to be considered valid MEDFORD, and therefore, the solution is obvious: the \texttt{desc} line can be re-conceptualized into a \texttt{name} line. Instead of being a freestyle annotation, the first line of a MEDFORD block is now considered to be its name.

Similar to a \texttt{name}, a \texttt{desc} can be any combination of ASCII characters and may include standard MEDFORD macros, and crucially, the \texttt{desc} can serve as a unique identifier for the purposes of cross-object referencing. In the above example, MEDFORD has the concept of a \texttt{@Species} object that is \texttt{name}d P. dam. With a name defined, we may now propose a syntax:
\begin{verbatim}
  @Species P.Dam
  @Species-Construction Pocillopora damicornis genome v1.0
  @Species-Reef New Caledonia Barrier Reef
\end{verbatim}




\subsection{Object Name Collisions}

This new adaptation of the \texttt{desc} syntax gives us names. We know that names have to be unique to avoid collisions, but to what degree? 

Consider the following:

\begin{verbatim}
  @Species P.Acuta
  @Species-Construction NCBI BioProject PRJNA812628

  @Photo P.Acuta
  @Photo-Type JPEG
\end{verbatim}

Here, the researcher describes a species of coral and includes a photo taken of it during the trip. Both use the same description or name, and as such would be keyed to the same identifier. By leveraging the block nature of MEDFORD, we can infer that these are different entities because they have different major tokens. When a researcher writes a MEDFORD block, that block is immediately associated with a major token. The major token defines the type of object being declared. As a result, we can not only contextually reuse names, but also provide some expectation of the type of metadata being referenced with the combination of a major token and name. As a result, we adjust the syntax to require the user to declare the major token:


\begin{verbatim}
  @Reef-Species @Species P.Acuta
\end{verbatim}

Accordingly, we have decided that MEDFORD names only have to be unique for blocks that share a major token. That is, if the user were to describe a \texttt{@Species}, its name must only be unique compared to the other \texttt{@Species}. This compromise is in line with reasonable naming guidance. Unique tag/name combinations will help users avoid misunderstandings and reduce errors downstream.




\subsection{Referencing Metadata from Other Sources}

While these new syntax features help expand the utility of MEDFORD, they are limited in their application by the requirement that the MEDFORD references be contained in the current document.

Consider the case where multiple research groups are studying the same set of coral species. We do not want to force the duplication of records by requiring everyone to copy the reference files to cite internally.
To account for such user stories, we extended our proposed reference functionality to add the capability to reference metadata stored in external MEDFORD files.




This requires two more additions to the MEDFORD syntax: the capability to declare that referenced metadata exists in a external file, and the capability to declare the location of an external file that will be referenced.


We have chosen to implement this via MEDFORD imports that ingest the entire file being referenced. While this may seem suboptimal from a computational resources standpoint, the boon to user experience more than makes up for it. Requiring a user to manually import every referenced metadata object would be incredibly detrimental to the efficacy of MEDFORD as a user-friendly system and increase the possibility of user error.

However, while solving one problem, this fix unfortunately creates a new issue: the potential of incongruent externally-referenced data. For example, suppose two separate MEDFORD files both define a \texttt{@Species P. dam}, but one references an older genome construction than the other. In that case, the MEDFORD language must have a way to deconvolute sources to avoid placing the responsibility on parsers to determine how to handle collisions.

To resolve this new issue, we leverage the new feature we have just introduced: every object in MEDFORD must have a \texttt{Name}, and therefore, the syntax to define an external MEDFORD file reference must also include a \texttt{Name}.

\subsection{External Reference Syntax}


\begin{verbatim}
  @Reef New Caledonia Barrier Reef
  @Reef-Species from CoralsMFD: @Species P.Dam
\end{verbatim}

This syntax aims to enhance usability by maximizing affordance and aligning expectations. For example, even without experience using Python or MEDFORD, a new reader would be likely to infer that \texttt{@Species P.Dam} lives in an object named \texttt{CoralsMFD}. This syntax choice enables users to nest provenance more deeply if needed, while maintaining parsable and predictable syntax. This design choice should increase users' ability to learn and reuse syntax across MEDFORD files quickly. Thus, MEDFORD is extensible across users and domains while maintaining a simple core set of syntax and rules~\cite{greenberg2012understanding,riley2017understanding}.

To formalize this relationship, we also include rules for codifying the namespace to file resolution. Consider:


\begin{verbatim}
  @Import CoralsMFD
  @Import-File ~/shared/corals_metadata.mfd
\end{verbatim}

 The example declares that MEDFORD metadata should be imported from the file \texttt{corals\_metadata.mfd} to be available for reference anywhere in the MEDFORD file using the nickname \texttt{CoralsMFD}.

\subsection{Redesigning Macro Syntax}


The reader may have noticed the awkwardness of retyping ``New Caledonia Barrier Reef'' repeatedly.
This serves as a motivating example of why MEDFORD was designed with a macro feature, so a shorter name could be expanded into something lengthy.
When the MEDFORD language was initially designed, macros used the programming-like syntax shown in Figure \ref{fig:macro_ex}. A macro is defined using a line that starts with the special characters \texttt{`@} directly followed by the macro's name. The name of a macro can only contain letters and numbers, not spaces or special characters. A space then follows it, and all further content until a newline is encountered. To use a macro, the macro header characters and the macro name are used in the payload of a line. Additionally, the macro name and header can be surrounded by curly braces, allowing researchers to prepend or append additional characters before or after the macro content. The importance of this is shown in figure \ref{fig:macro_curlies}.


\begin{figure}
  \begin{verbatim}
    `@Macro definition
    even more definition

    @Note `@Macro
    @Note Using the {`@Macro} a second time
  \end{verbatim}
  \caption{Example of defining and using a macro in v1.0 of the MEDFORD Language. There are two separate syntaxes for using a macro, one of which allows for additional text directly before or after the macro name, as the curly braces make the name of the macro clear.}
  \label{fig:macro_ex}
\end{figure}

\begin{figure}
  \begin{verbatim}
  `@CoralPhoto _pdam.png
  `@CoralPhotoDir ./data/photos/Jul27/

  @Photo 01{`@CoralPhoto}
  @Photo-File {`@CoralPhotoDir}01{`@CoralPhoto}

  @Photo 02{`@CoralPhoto}
  @Photo-File {`@CoralPhotoDir}02{`@CoralPhoto}

  # Resolved into:
  @Photo 01_pdam.png
  @Photo-File ./data/photos/Jul27/01_pdam.png

  @Photo 02_pdam.png
  @Photo-File ./data/photos/Jul27/02_pdam.png
  \end{verbatim}
  \caption{Example of macros being used to replace repetitive portions of data. Here, macros are being used to store consistent portions of filenames and file paths. Note how every macro use in this example is surrounded by curly braces; this deconvolutes the name of the macro from the characters after it, as otherwise any parser will have no means to identify the macro name \texttt{CoralPhotoDir} from the string \texttt{`@CoralPhotoDir01`@CoralPhoto}.}
  \label{fig:macro_curlies}
\end{figure}

However, this syntax encounters a major issue: MEDFORD allows for payloads to extend onto new lines. This is to support authorship, allowing researchers to separate their data onto newlines to better fit standard syntax (e.g., an address) or improve readability (e.g., large descriptive note blocks). Consider the case where two researchers share an institution, but have separate departments, as shown in Figure \ref{fig:macro_unclear_def}.

\begin{figure}
  \begin{verbatim}
    `@Institute University of Institutes
    42 Somerstreet
    Placeville, ST 02020

    @Contributor John Doe
    @Contributor-Affiliation Department of Anonymity
    `@Institute

    @Contributor Arthur Word
    @Contributor-Affiliation Department of Writing and Rhetoric
    `@Institute
  \end{verbatim}
  \caption{Example of the overlap between macro definition and macro usages. Since MEDFORD payloads can extend multiple lines, the researchers placed their \texttt{`@Institute} macro usages on new lines, much as an actual institutional address would be written. However, this creates lines that begin with the macro header characters \texttt{`@} -- therefore, creating macro definition lines. The parser cannot distinguish between these cases.}
  \label{fig:macro_unclear_def}
\end{figure}


Theoretically, this case could be resolved by forcing macros to always be referenced using the curly brace syntax described earlier. Still, the curly brace format was intended for more advanced users and is not something that we can expect all users to understand when first picking up MEDFORD. We cannot depend on the curly brace syntax, as as research shows how complex macro syntax can decrease adoption and learning among new users~\cite{crichton2024profiling}. Thus, this ambiguity reveals a necessity to revise the current syntax.

We are currently considering several options for the new macro syntax, placing an emphasis on separating the syntax of defining or invoking a macro. Several proposals include \texttt{>@} to define a macro and \texttt{<@} to invoke, or \texttt{`@} to define and \texttt{``@} to invoke. We aim to ensure that macro syntax is easy and quick to type, while not appearing too "code-like" and intimidating potential new users. We are planning user study experiments to determine which of these or other possible candidate solutions are most approachable and efficient.





\subsection{MEDFORD utilities: EXIF}

\begin{figure}[h!]
    \centering
    \includegraphics[width=0.75\linewidth]{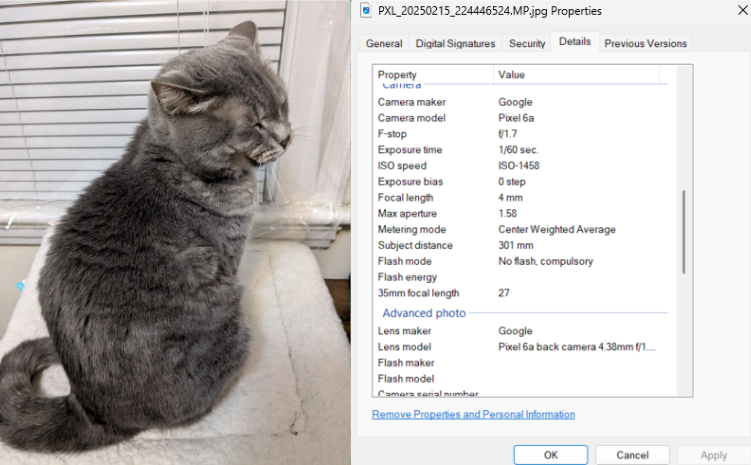}
    \caption[Example of a photograph and its EXIF metadata]{Example photograph alongside its EXIF metadata.}
    \label{fig:kiki}
\end{figure}

The external reference syntax detailed above can be adapted to provide even more functionality if we can extract the metadata already stored within native filetypes, such as a photograph date or GPS data if available, as shown in Figure~\ref{fig:kiki}. For example, photos taken by modern digital cameras automatically store metadata in a format called EXIF. This metadata could be linked to a MEDFORD entity using an \texttt{Import} block. Then, the import will reference the EXIF of the photo file, rather than requiring the researcher to re-enter the data themselves.
As we continue to implement updates to internal and external MEDFORD references, we will extend this capability to high-value file types.

In the interim, we have created a repository for small MEDFORD tools at \url{https://github.com/TuftsBCB/medfordTools}, and we have included a small script to create a MEDFORD file for an image file by parsing its EXIF data using the python \texttt{exif} package. These import tools will be added to future versions of the VS Code extension as import utilities.

\section{Conclusion and Future Work}

We have proposed additions to the MEDFORD language centered on connectivity and usability to ensure that intrinsically linked research data can be accurately documented in MEDFORD. These new syntax rules enable MEDFORD files to reference other MEDFORD files, allowing for data reuse and reducing transcription errors. To support these changes and prepare for increased user-generated data specifications, we have strengthened the data modeling and validation routines in MEDFORD and updated the VS Code extension and language server modules. We also plan an informal test to determine a new macro syntax, which we will then implement.

Continuing with the theme, our future work is also connectivity and usability focused. We are developing instructional materials and videos that will form the basis of building a community of MEDFORD users. Additionally, a variety of import plugins that allow the extraction or linking of metadata from data files (such as images and genomic sequencer files) could be useful. We also envision formalized user tests to determine how to best serve our scientific community, explore new major/minor tag requirements, and investigate how emerging technologies, such as intelligent agents, can utilize the context that tools like MEDFORD provide to enhance the user experience.
We are also investigating the development of a logging library for computational experiments (e.g., benchmarks) that will output directly to MEDFORD.

While our pilot application domain has been coral reef research, as that was the domain in which we were working when the need for something like MEDFORD became apparent, there are clear applications to numerous other domains.
In other areas of bioinformatics and biomedicine, the ability to easily collect metadata around experiments and sequencing runs will be of value.
In computational experiments that involve benchmarking and timing of algorithms and their implementations, the ability to easily collect results in MEDFORD will allow for more streamlined, better organized computational results.
In the realm of cybersecurity, a variety of formats exist for describing threats and intrusions~\cite{gao2022threatkg,abu2018cyber,barnum2012standardizing}.
However, lack of a unifying standard, or an interchange among them, could be addressed with MEDFORD.

Our goal is that these new features (namely, external references, flexible validation, and the improved VS Code extension) and future features (instructional materials, formal user testing, and other ideas to come) will help ensure that FAIR principles are easily applied to research data.

\section{Availability}

The MEDFORD parser and editor support are freely available at:
\begin{itemize}
\item \url{https://github.com/TuftsBCB/medford}
\item \url{https://github.com/TuftsBCB/medford-language-server}
\item \url{https://github.com/TuftsBCB/medford-vscode}
\end{itemize}

The development version of these tools discussed in this manuscript is available in the ``IJMSOrelease" branch of the repositories.


\section{Conflicts of Interest Declaration}

\textbf{All authors declare that they have no conflicts of interest.}





\bibliographystyle{abbrv}
\bibliography{referencesERC}
\end{document}